\documentclass[nofootinbib,reprint,amsmath,amssymb,prd,aps,superscriptaddress]{revtex4-1}
\usepackage{graphicx}
\usepackage[T1]{fontenc}
\usepackage{dcolumn}
\usepackage{xcolor,colortbl}
\usepackage{multirow}
\usepackage{comment}
\usepackage{enumitem}
\usepackage[caption=false]{subfig}
\usepackage{bm}

\usepackage{hyperref}
\hypersetup{colorlinks, linkcolor={teal},citecolor={blue},urlcolor={blue}}  

\newcommand{\dd}{{\rm d}}

\begin{document}

\title[Nonlocal gravity in a proper tetrad frame: traversable wormholes]{Nonlocal gravity in a proper tetrad frame: traversable wormholes}

\author{Rocco D'Agostino}
\email{rocco.dagostino@inaf.it}
\affiliation{INAF -- Osservatorio Astronomico di Roma, Via Frascati 33, 00078 Monte Porzio Catone, Italy}
\affiliation{INFN -- Sezione di Roma 1, P.le Aldo Moro 2, 00185 Roma, Italy}

\author{Vittorio~De~Falco}
\email{v.defalco@ssmeridionale.it}
\affiliation{Scuola Superiore Meridionale,  Largo San Marcellino 10, 80138 Napoli, Italy}
\affiliation{INFN -- Sezione di Napoli, Via Cintia, 80126 Napoli, Italy}

\begin{abstract}

We investigate the revised Deser-Woodard model of nonlocal gravity involving four auxiliary scalar fields, introduced to explain the standard cosmological background expansion history without fine-tuning issues. In particular, we simplify the complex field equations within a proper tetrad frame, thereby recasting the original system into a more tractable equivalent differential problem. We show that, by initially postulating the form of the $g_{tt}$ metric component, it is possible to reconstruct the distortion function of the gravitational model. We then describe a step-by-step procedure for solving the vacuum field equations in the case of a static and spherically symmetric spacetime. We apply our technique to find new traversable wormholes supported purely by gravity by employing either analytical, perturbative, or numerical methods. Furthermore, we demonstrate that the role of the nonlocal effects is analogous to that of exotic matter in general relativity, owing to their quantum nature. Finally, we discuss the main geometric properties of the obtained solutions. Our results present a feasible avenue for identifying novel compact objects while enhancing the comprehension of nonlocal gravitational theories.
\end{abstract}

\maketitle
\section{Introduction}

The quest for a comprehensive theory of gravity that bridges classical and quantum domains remains one of the most profound challenges in theoretical physics. While general relativity (GR) has been thoroughly validated via numerous experimental and observational tests \cite{Will:2014kxa,LIGOScientific:2016aoc,EventHorizonTelescope:2019dse}, it nonetheless faces theoretical limitations in both high and low-energy regimes \cite{Joyce:2014kja,Riess:2016jrr,Ishak:2018his,Barack:2018yly,DAgostino:2023cgx}. 
Although GR describes gravitation as the curvature of spacetime caused by mass and energy, it fails to provide a quantum description of gravity that would be consistent with the other fundamental forces \cite{Padmanabhan:2001ev,Ashtekar:2004eh}. These inconsistencies are particularly evident in extreme regimes, such as near black holes (BHs) or during the Planck era.

On the other hand, GR presents a different set of issues in the infrared regime, where the large-scale structure of the Universe is observed. The standard cosmological picture based on Einstein's gravity describes a Universe dominated by mysterious and largely unknown components \cite{SupernovaCosmologyProject:1998vns,Peebles:2002gy,Planck:2018vyg}. In particular, the cosmological constant, responsible for driving the accelerated expansion of the Universe at late times, conflicts with its interpretation as vacuum energy derived from Quantum Field Theory \cite{Weinberg:1988cp,Carroll:2000fy,DAgostino:2022fcx}.
This discrepancy suggests that the standard cosmological model may be incomplete, pointing to the need for alternative theoretical frameworks that could reconcile these differences and provide a more unified description of gravity, Quantum Mechanics, and the large-scale structure of the Universe \cite{Silvestri:2009hh,DeFelice:2010aj,Clifton:2011jh,Nojiri:2017ncd,Capozziello:2019cav,DAgostino:2019hvh,DAgostino:2022tdk}.

Among the various alternative theories explored in the last years to address the aforementioned challenges, a promising approach is to modify the gravitational sector by including nonlocal terms, thus encoding the influence of the whole spacetime. Nonlocal gravity models have shown their ability to reproduce inflationary dynamics, the formation of cosmic structures, and the dark energy features, but also to address BH and Big Bang singularities \cite{Arkani-Hamed:2002ukf,Barvinsky:2003kg,Calcagni:2007ru,Woodard:2018gfj,Nojiri:2007uq,Koivisto:2008xfa,Biswas:2011ar,Elizalde:2011su,Park:2012cp,Capozziello:2022rac,Deffayet:2024ciu}. 
Nonlocality features are typical of Quantum
Mechanics, making these models a potentially significant step toward a complete theory of Quantum Gravity.
Relaxing the classical locality principle provides a means to avoid the instabilities associated with higher-order derivative operators in ultraviolet extensions of GR, leading to renormalizable Lagrangians, and naturally incorporating nonlocal terms that emerge in loop corrections to effective Quantum Gravity actions \cite{Modesto:2011kw,Modesto:2017sdr,Knorr:2018kog,Buoninfante:2018mre}.

A notable example of nonlocal gravity theory is the Deser-Woodard model \cite{Deser:2007jk}, which involves the inverse of the d'Alembert operator acting on the Ricci scalar. This approach was originally proposed to reproduce the standard cosmological expansion history without fine-tuning issues. However, solar system experiments revealed that the model did not meet certain observational constraints due to the absence of a mechanism to screen nonlocal effects at short distances \cite{Belgacem:2018wtb}. To address these issues, the same authors refined their original model, leading to an improved version \cite{Deser:2019lmm}. This second framework was also considered for analyzing structure formation \cite{Ding:2019rlp,Jackson:2023faq}, bouncing cosmology \cite{Chen:2019wlu,Jackson:2021mgw}, and gravitational perturbations of the Schwarzschild BH \cite{Chen:2021pxd}. 

An interesting alternative to the Deser-Woodard model is the inclusion of a nonlocal term with a characteristic mass scale \cite{Jaccard:2013gla,Maggiore:2013mea,Maggiore:2014sia}. The latter could emerge dynamically from quantum gravity processes within the framework of GR, or through quantum corrections in models of massive gravity or theories involving extra dimensions.
In contrast with the Deser-Woodard scenario, in this case, the gravitational Lagrangian is predetermined, leaving the effective mass the only free parameter of the theory. Nonlocal models of this kind demonstrated theoretical consistency and exhibit an interesting cosmological phenomenology, providing a possible explanation of the origin of the dark sector while successfully fitting current cosmological data at both the background and linear perturbation levels \cite{Dirian:2014ara,Belgacem:2020pdz}. 

Even though significant advances have been made in nonlocal gravity, the search for astrophysical solutions in these frameworks often requires intricate analytical or numerical methods due to the increased complexity of the field equations compared to GR. 
Perturbative solutions to a static and spherically symmetric metric were obtained in \cite{Kehagias:2014sda,Kumar:2019uwi}, where it was shown that nonlocal infrared modifications of GR induced by a mass scale satisfy all solar system and laboratory experiment constraints. Additionally, other BH solutions were investigated via analytical and perturbative methods in \cite{Myung:2017qtc,Calcagni:2018pro,Buoninfante:2022ild}. However, solving the equations of motion in these models remains notoriously difficult, complicating the description of compact object configurations. 

The aim of this paper is to investigate wormhole (WH) solutions within the framework of the improved Deser-Woodard model. 
WHs are exotic compact objects characterized by the requirement that the spacetime remains smooth everywhere. These structures connect two asymptotically flat regions through a slender bridge or throat, free from event horizons and central singularities \cite{Visser:1995cc}. For stability and traversability within the framework of GR, WHs require the presence of exotic matter, which involves mechanisms that violate the standard energy conditions \cite{Morris:1988cz}. Research on WHs can be broadly divided into two main areas: (1) formulating new WH solutions within the framework of GR \cite{Armendariz-Picon:2002gjc,Bronnikov:2013coa,Garattini:2019ivd,Bambi:2021qfo,Konoplya:2021hsm} or alternative gravity theories, utilizing either exotic stress-energy tensors \cite{Ebrahimi:2009rv,Boehmer:2012uyw,Harko:2013yb,Bahamonde:2016jqq,Mustafa:2021vqz}, purely gravitational topological configurations \cite{DeFalco:2023twb}, or matter fields that adhere to the energy conditions \cite{Eiroa:2008hv,Lobo:2009ip,Capozziello:2012hr}; (2) devising novel astrophysical strategies to possibly detect observational evidence of WHs, employing techniques in the X-ray domain \cite{Ohgami:2015nra,Paul:2019trt,Dai:2019mse,DeFalco:2020afv,DeFalco:2021klh,DeFalco:2021ksd,DeFalco:2023kqy} and gravitational-wave astronomy \cite{Hong:2021qnk,Konoplya:2016hmd,Cardoso:2016rao,DeSimone:2025sgu}.

This work follows the first research line and the results of the previous study \cite{DAgostino:2025wgl}, inspired by the pioneering paper of Morris and Thorne \cite{Morris:1988cz} and the recent advancements in nonlocal gravity cosmology \cite{Capozziello:2023ccw}. Here, we employ a strategy to rewrite the Deser-Woodard field equations in a proper tetrad frame, simplifying their structure, while preserving the essential physics. Our methodology is also capable of determining three solutions of traversable static and spherically symmetric WHs.

The paper is organized as follows: in Sec.~\ref{sec:nonlocal-theory}, we briefly review the fundamentals of the Deser-Woodard nonlocal theory; in Sec. \ref{sec:localizing-strategy}, we present our approach based on writing the nonlocal field equations in an appropriate tetrad frame and proposing a strategy to solve them; in Sec.~\ref{sec:WH-solutions}, we determine static and spherical WH solutions; in Sec.~\ref{sec:WH-properties}, we discuss the main geometrical properties of the obtained solutions; in Sec.~\ref{sec:end}, we draw our conclusions and outline the future perspectives of our work.

Throughout this paper, we adopt units of $c=\hbar=1$. The flat metric is indicated by $\eta_{\alpha \beta }={\rm diag}(-1,1,1,1)$.

\section{Deser-Woodard nonlocal gravity}
\label{sec:nonlocal-theory}
We consider the Deser-Woodard model of nonlocal gravity, whose action is defined as  \cite{Deser:2019lmm}
\begin{equation} \label{eq:nonlocal-action}
S=\dfrac{1}{16\pi G}\int \dd^4 x \sqrt{-g}\, R\left[1+f(Y)\right],
\end{equation}
where $g$ is the detrminant of the metric tensor $g_{\mu \nu}$, and $R$ is the Ricci scalar. 
Here, $f(Y)$ is the so-called distortion function, defined in terms of differential equations involving the following two auxiliary scalar fields:
\begin{align}
    \Box X&=R\,, \label{eq:X} \\ 
    \Box Y&=g^{\mu\nu}\partial_\mu X\partial_\nu X\,, \label{eq:Y}
\end{align}
where $\Box\equiv \nabla_\mu \nabla^\mu$ is the relativistic d’Alembert operator, which can be defined on a function $u$ as
\begin{equation} 
    \Box u\equiv\frac{1}{\sqrt{-g}}\partial_\alpha\left[\sqrt{-g}\,\partial^\alpha u\right].
    \label{eq:local action}
\end{equation}
The action \eqref{eq:nonlocal-action} can be recast in terms of two auxiliary scalar fields $U$ and $V$, treated both as Lagrange multipliers, in the localized form
\begin{equation}\label{action_localized}
S=\dfrac{1}{16\pi G}\int \dd^4x \sqrt{-g}\,\Big\{R\left[1+U+f(Y)\right]+g^{\mu\nu}\mathcal{K}_{\mu\nu}\Big\}\,,
\end{equation}
where the following tensor has been introduced:
\begin{equation}\label{eq:B}
\mathcal{K}_{\mu\nu}:= \partial_\mu X\partial_\nu U +\partial_\mu Y \partial_\nu V+V\partial_\mu X \partial_\nu X \,.
\end{equation}
The differential equations governing the dynamics of the fields $U$ and $V$ can be determined by varying the action \eqref{action_localized} with respect to $X$ and $Y$, respectively, so to obtain
\begin{align}
    \Box U&=-2\nabla_\mu (V\nabla^\mu X)\,, \label{eq:U} \\
    \Box V&=R\frac{\dd f}{\dd Y}\,.\label{eq:V}
\end{align}
It is worth emphasizing that, within this framework, the scalar fields $X$, $Y$, $U$, and $V$ are independent and all satisfy Klein-Gordon equations, while the action \eqref{eq:local action} is considered to be local. Moreover, to avoid the presence of ghost-like instabilities, all auxiliary scalar fields must obey retarded boundary conditions, vanishing along with their first-time derivatives at the initial value surface \cite{Deser:2013uya}.

The vacuum field equations can be then obtained by varying the action \eqref{eq:local action} with respect to $g^{\mu\nu}$ \cite{Deser:2019lmm}:
\begin{align}
&\left(G_{\mu\nu}+g_{\mu\nu}\Box-\nabla_\mu \nabla_\nu\right)\left[1+U+f(Y)\right]+\mathcal{K}_{(\mu\nu)}\notag  \\
&-\frac{1}{2}g_{\mu\nu}g^{\alpha\beta}\mathcal{K}_{\alpha\beta} =0\,,
\label{eq:FE}
\end{align}
where 
$\mathcal{K}_{(\mu\nu)}\equiv (\mathcal{K}_{\mu\nu}+\mathcal{K}_{\nu\mu})/2$.

In the following section, we describe the strategy devised in \cite{DAgostino:2025wgl} to recast Eqs.~\eqref{eq:FE} in a suitable tetrad frame, showing how it is possible to simplify and solve the aforementioned differential problem.  

\section{Nonlocal gravity in the proper tetrad frame}
\label{sec:localizing-strategy}

Let us start from a generic static and spherically symmetric metric, written in spherical-like coordinates $(t,r,\theta,\varphi)$, whose line element reads as
\begin{equation}
\dd s^2=g_{tt}(r) \dd t^2+g_{rr}(r)\dd r^2 +r^2(\dd\theta^2+\sin^2\theta\, \dd\varphi^2)\,,
    \label{eq:general-metric}
\end{equation}
where $g_{tt}$ and $g_{rr}$ are unknown functions of the radial coordinate, $r$. 

We thus consider the orthonormal tetrad field associated with a static observer located at infinity, $\left\{{\bf e}_t\,, {\bf e}_r\,,{\bf e}_\theta\,,{\bf e}_\varphi\right\}=\left\{\partial_t\,,\partial_r\,,\partial_\theta,\partial_\varphi\right\}$. In particular, for a static observer in the spacetime \eqref{eq:general-metric}, we consider the tetrad frame
\begin{align}
{\bf e}_{\hat t} = \frac{{\bf e}_t}{\sqrt{-g_{tt}}},\ 
{\bf e}_{\hat r}= \frac{{\bf e}_r}{\sqrt{g_{rr}}},\
{\bf e}_{\hat \theta} =\frac{{\bf e}_\theta}{r},\ 
{\bf e}_{\hat \varphi}=\frac{{\bf e}_\varphi}{r\sin\theta},
\end{align}
such that $g_{\hat\mu\hat\nu}=e_{\hat \mu}^\alpha e_{\hat \nu}^\beta g_{\alpha\beta}\equiv\eta_{\mu\nu}$, where
\begin{equation}\label{eq:general-tetrad}
    e^{\hat \alpha}_{\mu}:=\text{diag}\left(\frac{1}{\sqrt{-g_{tt}}},\frac{1}{\sqrt{g_{rr}}},\frac{1}{r},\frac{1}{r\sin\theta}\right).
\end{equation}
Therefore, the Riemann tensor transforms as
\begin{equation}
R^{\hat \alpha}{}_{{\hat \beta}{\hat \gamma}{\hat \delta}} = e^{\hat \alpha}_{\mu} e_{\hat \beta}^{\nu} e_{\hat \gamma}^{\rho} e_{\hat \delta}^{\sigma} R^\mu{}_{\nu \rho \sigma}\,,
\end{equation}
and the Ricci tensor and scalar are given by, respectively,
\begin{equation}
    R_{\hat\mu\hat\nu}=R^{\hat\alpha}_{\ \hat\mu\hat\alpha\hat\nu}\,,\quad  R=\eta^{\hat\mu\hat\nu}R_{\hat\mu\hat\nu}\,.
\end{equation}
Hence, the Einstein tensor reads
\begin{equation}
    G_{\hat\mu\hat\nu}=R_{\hat\mu\hat\nu}-\frac{1}{2}\eta_{\mu\nu} R\,.
\end{equation}
Then, Eq.~\eqref{eq:FE} takes the form\footnote{Note that the  D'Alembert operator is invariant under tetrad transformations.} 
\begin{align}
&\left(G_{\hat\mu\hat\nu}+\eta_{\mu\nu}\Box-\nabla_{\hat\mu} \nabla_{\hat\nu}\right)W
+\mathcal{K}_{(\hat\mu\hat\nu)}-\frac{1}{2}\eta_{\mu\nu}\eta^{\alpha\beta}\mathcal{K}_{\hat\alpha\hat\beta} =0\,,
\label{eq:Local_FE}
\end{align}
where $\nabla_{\hat\mu} \nabla_{\hat\nu}=e^\alpha_{\hat\mu}e^\beta_{\hat\nu}\nabla_\alpha \nabla_\beta$, $\mathcal{K}_{\hat\mu\hat\nu}=e^\alpha_{\hat\mu}e^\beta_{\hat\nu}\mathcal{K}_{\alpha\beta}$, and
\begin{equation}\label{eq:W}
W(r):= 1+U(r)+f(Y(r))\,.   
\end{equation}

The non-vanishing components of Eq.~\eqref{eq:Local_FE} are \cite{DAgostino:2025wgl}
\begin{subequations}
\begin{align}
(G_{\hat t\hat t}-\Box -\nabla_{\hat t}\nabla_{\hat t})W+\frac{1}{2}\mathcal{K}_{\hat r\hat r}&=0,\label{eq:FE_LOC_t}\\    
(G_{\hat r\hat r}+\Box -\nabla_{\hat r}\nabla_{\hat r})W+\frac{1}{2}\mathcal{K}_{\hat r\hat r}&=0,\label{eq:FE_LOC_r}\\ 
(G_{\hat \varphi\hat \varphi}+\Box -\nabla_{\hat \varphi}\nabla_{\hat \varphi})W-\frac{1}{2}\mathcal{K}_{\hat r\hat r}&=0.\label{eq:FE_LOC_phi}    
\end{align}
\end{subequations}
Combining Eqs.~\eqref{eq:FE_LOC_t}, \eqref{eq:FE_LOC_r}, and  \eqref{eq:FE_LOC_phi}, we find the following independent equations:
\begin{subequations}\label{eq:LOC_SUM}
\begin{align}
    &(G_{\hat t\hat t}+G_{\hat\varphi\hat\varphi})W=(\nabla_{\hat t}\nabla_{\hat t}+\nabla_{\hat\varphi}\nabla_{\hat\varphi})W\,,\label{eq:LOC_SUM1} \\
    &(G_{\hat r\hat r}+G_{\hat\varphi\hat\varphi})W+2\Box W=(\nabla_{\hat r}\nabla_{\hat r}+\nabla_{\hat\varphi}\nabla_{\hat\varphi})W\,.\label{eq:LOC_SUM2}
\end{align}
\end{subequations}
We note that the above equations are easier to handle compared to Eq.~\eqref{eq:FE}, see Appendix \ref{sec:full}. 

In order to determine the radial behavior of the auxiliary fields $\{X,Y,U,V\}$ and, consequently, obtain the distortion function $f(Y)$, we start by postulating the functional form of $\Phi(r)$ and then solving the system \eqref{eq:LOC_SUM1} and \eqref{eq:LOC_SUM2} for $W(r)$ and $b(r)$. The integration constants can be determined by imposing appropriate boundary conditions, depending on the problem under study.

The metric tensor permits to determine the Ricci scalar, which can be used to obtain $X(r)$ from Eq.~\eqref{eq:X}. Substituting the solution to the latter into Eq.~\eqref{eq:Y} yields
\begin{equation}\label{eq:Y_1}
    \Box Y=g^{rr}(X')^2\,,
\end{equation}
which will provide us with $Y(r)$. Here, the prime denotes the derivative with respect to $r$.

Moreover, from Eq.~\eqref{eq:W} we have
\begin{equation}\label{eq:f(r)}
f(r)=W(r)-U(r)-1\,,
\end{equation} 
so that we can write  
\begin{equation}
\frac{\dd f}{\dd Y}=\frac{f'}{Y'}=\frac{W'-U'}{Y'}.
\label{eq:f'(Y)}
\end{equation}
Additionally, we can rearrange Eq.~\eqref{eq:U} as \cite{Capozziello:2023ccw}
\begin{equation}
    U'=-2VX'\,.
    \label{eq:U_1}
\end{equation}
With the help of Eqs.~\eqref{eq:U_1} and \eqref{eq:f'(Y)}, Eq.~\eqref{eq:V} becomes
\begin{equation}\label{eq:V_1}
\Box V=\left(\frac{W'+2VX'}{Y'}\right)R\,. 
\end{equation}
Solving the latter will allow us to determine $V(r)$ and, thus, the solution to Eq. \eqref{eq:U_1} will provide $U(r)$.

Finally, Eq.~\eqref{eq:f(r)} can easily give $f(r)$. Then, inverting the function $Y(r)$, one gets $r(Y)$, which can be plugged into $f(r)$ to obtain the distortion function $f(Y)$.

\section{Static and spherically symmetric wormhole solutions}
\label{sec:WH-solutions}

We shall look here for WH solutions arising from the nonlocal gravity theory  \eqref{action_localized}. For this purpose, we consider the static and spherically symmetric spacetime \cite{Morris:1988cz}
\begin{equation}
    \dd s^2=-e^{2\Phi(r)} \dd t^2+\frac{\dd r^2}{1-\frac{b(r)}{r}} +r^2(\dd\theta^2+\sin^2\theta\, \dd\varphi^2)\,,
    \label{eq:WH-metric}
\end{equation}
where $\Phi(r)$ and $b(r)$ are known as the redshift and shape functions, respectively. The radial coordinate belongs to the domain $ \mathcal{D}: (-\infty,-r_0]\cup[r_0,\infty)$, where the positive and negative values of $r$ refer to the two symmetric universes joined by the WH throat represented by $r_0>0$. Due to the spherical symmetry hypothesis, we can set $\theta=\pi/2$ without loss of generality.

In this framework, the following conditions must hold:
\begin{enumerate}[label=(\roman*)]
    \item The metric must be \emph{asymptotic flat} in the two universes, namely 
\begin{equation}
\lim_{r \to \pm\infty}\Phi(r)=0\,,\qquad \lim_{r \to \pm\infty}\frac{b(r)}{r}=0\,.
\end{equation}
    \item $\Phi(r)$ and $b(r)$ are smooth and finite functions in $\mathcal{D}$, to avoid horizons and essential singularities. Furthermore, $\Phi(r)$ and $b(r)/r$ are monotonically increasing and decreasing functions, respectively.
\item We require $b(r)\le r$ and $b(r_0)=r_0$.
\item In order to have a stable and traversable WH, the \emph{flaring out condition} must hold:
\begin{equation}\label{eq:FOC}
    b(r)-rb’(r) < 1\,, \ \text{near}\ r=r_0\,.
\end{equation}
\end{enumerate}
The tetrad field \eqref{eq:general-tetrad} applied to the metric \eqref{eq:WH-metric} reads
\begin{align}
{\bf e}_{\hat t} &= e^{-\Phi(r)}{\bf e}_t \,, & {\bf e}_{\hat r} &= \sqrt{1-\frac{b(r)}{r}}\,{\bf e}_r \,, \notag \\
{\bf e}_{\hat \theta} &= \frac{1}{r}\,{\bf e}_\theta \,, & {\bf e}_{\hat \varphi} &= \frac{1}{r\sin\theta}\,{\bf e}_\varphi \,.
\end{align}
The non-vanishing components of $G_{\hat\mu\hat\nu}$ are \cite{Morris:1988cz}:
\begin{subequations}
\begin{align}
G_{\hat t\hat t}&=\frac{b'}{r^2}\,, \\
G_{\hat r\hat r}&=-\frac{b}{r^3}+2\left(1-\frac{b}{r}\right)\frac{\Phi'}{r}\,, \\
G_{\hat\theta\hat\theta}&=\left(1-\frac{b}{r}\right)\left[\Phi''+(\Phi')^2-\frac{b'r-b}{2r(r-b)}\Phi'+\frac{\Phi'}{r}\right.\notag\\
&\left. \hspace{2cm} -\frac{b'r-b}{2r^2(r-b)}\right], \\
G_{\hat\varphi\hat\varphi}&=G_{\hat\theta\hat\theta}\,.
\end{align}
\end{subequations}
The Ricci curvature scalar reads as
\begin{equation}\label{eq:R}
R=\frac{b' \left(r \Phi'+2\right)+(3 b-4 r) \Phi'}{r^2}-2\left(1-\frac{b}{r}\right) \left[\Phi''+(\Phi')^2\right].
\end{equation}
It appears then evident that the problem under consideration is quite complex from an analytical point of view. However, we show here how to find three WH solutions via analytical, perturbative, and numerical methods. Although they look simple in form, they require considerable effort to be determined.

Under the spacetime \eqref{eq:WH-metric}, the field equations \eqref{eq:LOC_SUM} read 
\begin{subequations}
\begin{align}
&r W b' \left(1-r \Phi '\right)+2 r^2 W (r-b) \Phi ''-2 r (b-r) W' \left(r \Phi '-1\right)\notag\\
&+W \left(r \Phi '+1\right) \left[2 r (r-b) \Phi '+b\right]=0\,,\label{eq:WH-FE1}\\
&r \Big\{2 r (r-b) \left(W''+W \Phi ''\right)-b' \left(r W'+r W \Phi '+W\right)\notag\\
&+\Phi ' \left[4 r (r-b) W'-5 b W+6 r W\right]+(6 r-5 b) W'\notag\\
&+2 r W (r-b) \left(\Phi '\right)^2\Big\}-b W=0\,.\label{eq:WH-FE2}
\end{align}    
\end{subequations}
We note that Eqs.~\eqref{eq:WH-FE1} and \eqref{eq:WH-FE2} involve up to the second derivative of $\Phi(r)$ and the first derivative in $b(r)$. For this reason, it is more reasonable to specify the functional form of $\Phi(r)$, as this approach is more likely to yield an analytical solution for $b(r)$. Conversely, approaching the problem in the opposite way is more challenging when attempting to achieve analytical objectives.
 
Requiring the asymptotic flatness implies that  $R\to0$ for $r\to\infty$. Consequently, all auxiliary fields $\{X,Y,U,V\}$ must also vanish at $r\rightarrow \infty$. Furthermore, to recover GR at infinity, namely $f(Y)\to0$, we must require that $W(r)\rightarrow 1$ for $r\rightarrow \infty$.
In the following calculations, we name the integration constants arising from the solutions of the above differential equations using a lowercase letter matching the corresponding scalar field (e.g., $x_1$ is related to $X(r)$, $y_1$ to $Y(r)$ and so forth). 
These constants are chosen such that the auxiliary scalar fields vanish for very large radii, thereby recovering the GR behavior. Additionally, to simplify the notation, we set $r_0=1$.

\subsection{$\Phi(r)=\Phi_0$}
\label{sec:I-case}

As a first attempt, let us consider $\Phi(r)=\Phi_0=\text{const}$\footnote{This assumption does not compromise the asymptotical flatness, which can still be achieved through the redefinition $\dd \tilde{t}=e^{\Phi_0}\dd t$.}. In this case, we look for a shape function of the form $b(r)=1/r^n$, where $n>0$ is a constant to be determined. 

From Eq. \eqref{eq:WH-FE1} we find
\begin{equation}\label{eq:SolW1}
W(r)=w_1 \exp \left\{\int\frac{b(r)+r b'(r)}{2 r [r-b(r)]}\dd r\right\}, 
\end{equation}
where $w_1$ is an integration constant. Plugging this solution in Eq.~\eqref{eq:WH-FE2} yields $n=1$. Thus, we have 
\begin{equation}
    b(r)=\frac{1}{r}\,,
\end{equation}
which can be inserted back into Eq.~\eqref{eq:SolW1} to find
\begin{equation}\label{eq:W_sol1}
    W(r)=1\,.
\end{equation}
In this case, Eq.~\eqref{eq:R} gives the Ricci scalar as
\begin{equation}
    R(r)=-\frac{2}{r^4}\,,
\end{equation}
which can be used in Eq.~\eqref{eq:X} to obtain
\begin{equation} \label{eq:X_sol1}
    X(r)=\frac{\pi ^2}{4}-\lambda_1^2\,,         
\end{equation}
where $\lambda_1\equiv\arctan\left(\sqrt{r^2-1}\right)$.

Moreover, from Eq.~\eqref{eq:Y_1} we get
\begin{equation}
    Y(r)=\frac{1}{3}\left( \lambda_1^4-\frac{\pi ^4}{16}\right).
    \label{eq:Y_sol1}
\end{equation}
The resolution of Eq.~\eqref{eq:V_1} gives
\begin{equation}\label{eq:V_sol1}
    V(r)=v_1\left(\lambda_1^3-\frac{\pi ^5}{32 \lambda_1^2}\right),
\end{equation}
with $v_1$ being an integration constant. Then, the solution to Eq.~\eqref{eq:U_1} is given as 
\begin{align}\label{eq:U_sol1}
    U(r)&=\frac{v_1}{5}\Bigg\{4\lambda_1^5-\frac{\pi^5}{8}\left[1+\ln \left(\frac{32 \lambda_1^5}{\pi ^5}\right)\right]\Bigg\}.
\end{align}
The value of the constant $v_1$ can be determined by requiring that Eqs.~\eqref{eq:FE_LOC_t}--\eqref{eq:FE_LOC_phi}, as well as Eq.~\eqref{eq:FE}, are satisfied for the metric tensor and auxiliary fields obtained above. In doing so, one finds that 
\begin{equation}
    v_1=\frac{48}{5 \pi ^5}\,.
\end{equation}

Finally, from Eq.~\eqref{eq:f(r)}, we have
\begin{align}\label{eq:f(r)_sol1}
f(r)&=\frac{6}{25}\Bigg\{\left[1+\ln \left(\frac{32 \lambda_1^5}{\pi ^5}\right)\right]-\frac{32}{\pi^5}\lambda_1^5\Bigg\}.
\end{align}
Inverting Eq.~\eqref{eq:Y_sol1} yields
\begin{equation}\label{eq:r(Y)_sol1}
r(Y)= \sec \left[\frac{1}{2}\left( 48 Y+\pi ^4\right)^{1/4}\right],    
\end{equation}
where $Y\in[-\pi^4/48,0]$. Equation \eqref{eq:r(Y)_sol1} can be then substituted in Eq.~\eqref{eq:f(r)_sol1} to obtain the distortion function $f(Y)$ defining the nonlocal gravity theory:
\begin{align}
    f(Y)&=\frac{6}{25} \left\{1-\left[\frac{48Y}{\pi ^4}+1\right]^{\frac{5}{4}}+ \ln\left[16 \left(\frac{48Y}{\pi ^4}+1\right)^{\frac{5}{4}}\right]\right\}.
\end{align}

\begin{figure*}
\centering
    \begin{minipage}{0.47\textwidth}
    \centering
    \includegraphics[width=3.2in]{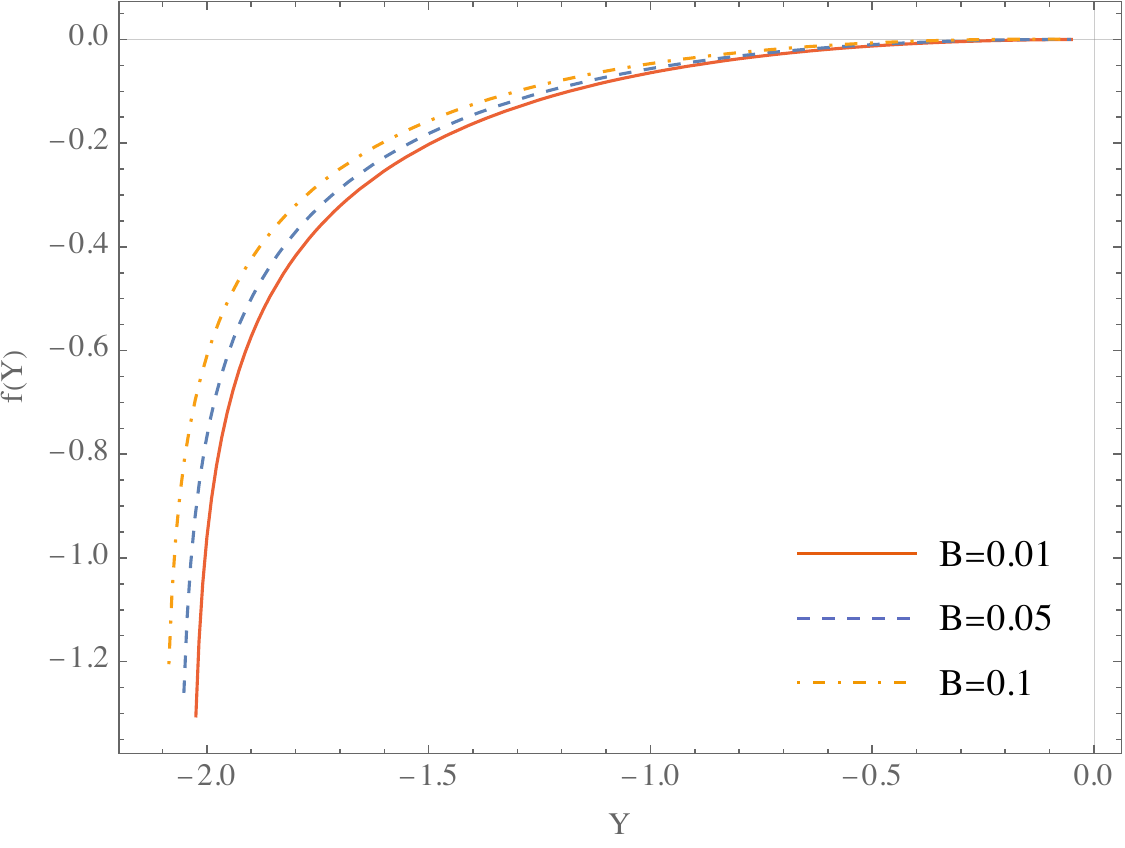}
    \caption{Behavior of the distortion function (see Eq.~\eqref{eq:f(r)}) of the WH2 solution for different values of the parameter $B$. The results are based on the analytical expressions for the auxiliary scalar fields given in Eqs.~\eqref{eq:W_IIcase_sol},  \eqref{eq:U1_IIsol} and \eqref{eq:f1_IIsol}. The conversion $f(r)\to f(Y)$ is performed using Eq.~\eqref{eq:r(Y)_sol1}.}
    \label{fig:fY}
    \end{minipage}  
    \hfill
    \begin{minipage}{0.47\textwidth}
    \centering
    \includegraphics[width=3.2in]{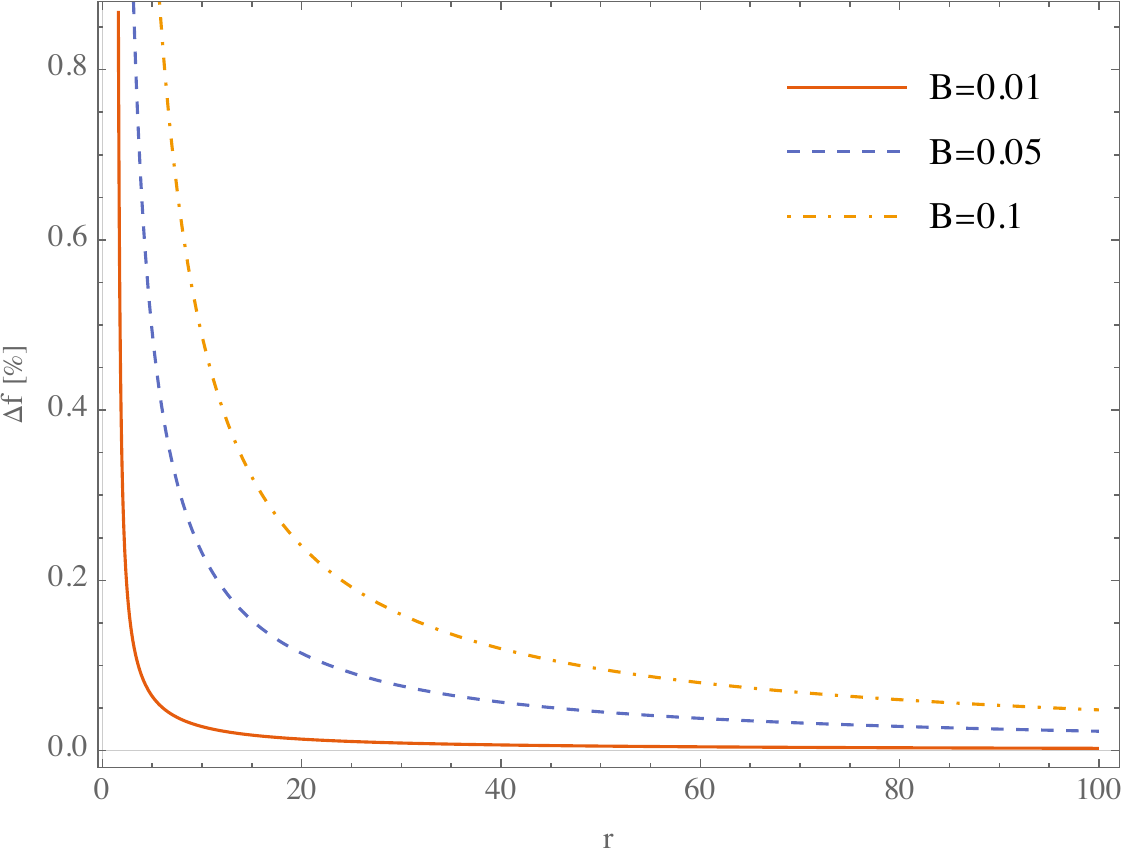}
    \caption{Absolute difference between the  distortion functions of the WH3 and WH2 solutions for different values of the parameter $B$. 
    The difference decreases for smaller  value of $B$, reflecting the accuracy of the linear approximation for $\Phi(r)$: $\frac{1}{2}(1-\frac{B}{r})\simeq -\frac{B}{2r}$ (see discussion in Sec.~\ref{sec:II-case}).} 
    \label{fig:deltaf}
    \end{minipage}
\end{figure*}

\subsection{$\Phi(r)= \frac{1}{2}\ln\left(1-\frac{B}{r}\right)$}
\label{sec:II-case}
As second analysis, we consider the redshift function $\Phi(r)=\frac{1}{2}\ln(1-\frac{B}{r})$. Specifically, in the perturbative regime for $0< B\ll 1$, with $r_0>B$ \cite{Morris:1988cz}, at the linear order, we have $\Phi(r)\simeq -\frac{B}{2r}$. This case is particularly interesting as it allows us to obtain analytical solutions and represents a generalization of the model investigated in Sec.~\ref{sec:I-case}, which is readily recovered when $B\rightarrow 0$.
It is important to note that the parameter $B$, introduced by the redshift function $\Phi(r)$, is also associated with the shape of the WH, influencing its size and geometry. This parameter does not directly govern deviations from GR, which, in our case, arise from the nonlocal modifications to the Einstein field equations. Thus, while $B$ plays a crucial role in determining the WH structure, it does not inherently indicate a departure from standard gravity.

At the linear order in $B$, from Eq.~\eqref{eq:WH-FE1} we have
\begin{equation}
   W(r)= w_1 \exp \left\{\int\frac{1}{2 (r-b)}\left[b'+\frac{(B-2 r) (B-b)}{(3 B-2 r) (B-r)}\dd r\right]\right\},
   \label{eq:W_IIcase}
\end{equation}
where $w_1$ is an integration constant. Substituting Eq.~\eqref{eq:W_IIcase} into Eq.~\eqref{eq:WH-FE2}, expanding at first order in $B$, and considering $b(r)=\frac{1}{r}+B\, h(r)$, we obtain 
\begin{equation}\label{eq:b_sol2}
b(r)= \frac{1}{r}+B\left(\frac{1-r+\sqrt{r^2-1}-\lambda_1}{r}\right).
\end{equation}

Hence, Eq.~\eqref{eq:W_IIcase} provides us with
\begin{equation}
    W(r)=1+B\left(\frac{1}{r}-\lambda_2\right),
    \label{eq:W_IIcase_sol}
\end{equation}
where $\lambda_2\equiv\arctan\left(r-\sqrt{r^2-1}\right)$.

The Ricci scalar \eqref{eq:R} reads 
\begin{equation}
R(r)=-\frac{2}{r^4}+B\left(\frac{2 r \lambda_1-2 r-1}{r^5}\right). 
\end{equation}
Therefore, one can solve Eq.~\eqref{eq:X}  for $X(r)=X_0(r)+B X_1(r)$, where $X_0(r)$ is the zeroth-order solution given in Eq.~\eqref{eq:X_sol1}, while 
\begin{align}\label{eq:X_sol2}
X_1(r)&=\frac{2 \lambda_1^5}{15}+\frac{8}{r}+\frac{\pi ^4}{24}-\frac{\pi ^3}{6}-\frac{\pi ^5}{240}-2 \pi\notag\\
&-\frac{2 \lambda_1^3 \left(\lambda_3 \sqrt{r^2-1} r-2 r^2+r+1\right)}{3 r \sqrt{r^2-1}}+\frac{4 \lambda_1 \sqrt{r^2-1}}{r}\notag\\
&+\frac{2}{3} \lambda_1^4 \left(\frac{1}{\sqrt{r^2-1}}-1\right)\,,
\end{align}
where $\lambda_3\equiv\arctan \left(1/\sqrt{r^2-1}\right)$.

Considering Eq.~\eqref{eq:Y_1} and using the same strategy devised for $X(r)$, we write $Y(r)=Y_0(r)+BY_1(r)$, where $Y_0(r)$ is given by Eq.~\eqref{eq:Y_sol1} and 
\begin{align}\label{eq:Y_sol2}
Y_1(r)&=-\frac{2}{3 r \sqrt{r^2-1}} \left(1+r-2 r^2+\sqrt{r^2-1} r \lambda_3\right) \lambda_1^3\notag\\
&+\frac{2}{3} \left(\frac{1}{\sqrt{r^2-1}}-1\right) \lambda_1^4+\frac{2}{15} \lambda_1^5+\frac{8}{r}+\frac{\pi ^4}{24}-\frac{\pi ^3}{6}\notag\\
&-\frac{\pi ^5}{240}-2 \pi+\frac{4\lambda_1}{r}\sqrt{r^2-1}\,.
\end{align}

Similarly, we look for a solution of Eq.~\eqref{eq:V_1} under the form $V(r)=V_0(r)+B\,V_1(r)$, where $V_0(r)$ is given by Eq. \eqref{eq:V_sol1}. In this case, we report the lengthy expression of $V_1(r)$ in Appendix~\ref{sec:appendix} (see Eq.~\eqref{eq:V1_IIsol}), depending on the integration constant $v_2$. Substituting the solution for $V(r)$ into Eq.~\eqref{eq:U_1}, we obtain the analytical expression for $U(r)$ as reported in Eq.~\eqref{eq:U1_IIsol}.
Also in this case, the value of the constant $v_2$ can be found by imposing that Eqs.~\eqref{eq:FE_LOC_t}--\eqref{eq:FE_LOC_phi} (as well as also Eqs. \eqref{eq:FE}) are fulfilled for the metric tensor and auxiliary fields now determined. Therefore, we have
\begin{align}
    v_2&=\frac{1}{\pi^2}\left(\frac{288}{\pi ^6}-\frac{84}{\pi ^4}+\frac{672}{25 \pi ^3}-\frac{279}{100 \pi ^2}-\frac{7}{160}\right)-\frac{7 \text{Si}\left(\frac{\pi }{2}\right)}{640}\,,
\end{align}
where $\text{Si}(z)\equiv \int_0^z\frac{\sin x}{x} \dd x$\,.

One can finally find $f(r)$ from Eq.~\eqref{eq:f(r)} (see Eq.~\eqref{eq:f1_IIsol}). Unfortunately, determining $f(Y)$ analytically is not possible because Eq.~\eqref{eq:Y_sol2} cannot be inverted. Therefore, we employ an approximation strategy by substituting Eq.~\eqref{eq:r(Y)_sol1} into $f(r)$ (see Eqs.~\eqref{eq:f(r)_sol1} and \eqref{eq:f1_IIsol}). For $B=0.01$, we find that the relative error with respect to the numerical outcome goes from $\sim 40\%$ for $r\approx 1.1$ to $\sim 1\%$ for $r\gtrsim1.08$. This means our approximation works well beyond a close region around the WH throat. 

Finally, it is interesting to analyze the full solution  $\Phi(r)=\frac{1}{2}\ln\left(1-\frac{B}{r}\right)$ for a generic value of $B$. In this case, the $g_{rr}$ metric cannot be calculated analytically, and one is forced to resort to numerical integration with boundary conditions $b(1)=1$ and $b(\infty)=0$. This implies that all auxiliary scalar fields, together with the distortion function, must also be calculated numerically. 
The boundary conditions are $X(\infty)=Y(\infty)=V(\infty)=U(\infty)=0$, $X'(\infty)=Y'(\infty)=V'(\infty)=0$ and $W(\infty)=1$. Finally, one can obtain the distortion function by considering the parametric plot $[Y(r),f(r)]$. It is worth noting that both the perturbative and numerical solutions fulfill all the properties listed in Sec.~\ref{sec:WH-solutions}.

In what follows, we refer to the solution of Sec.~\ref{sec:I-case} as WH1, whereas to the perturbative and numerical solutions of Sec.~\ref{sec:II-case} as WH2 and WH3, respectively. 

In Fig.~\ref{fig:fY}, we display the behavior of the distortion function of the WH2 solution for different values of $B$. We notice that the difference between the curves becomes smaller as $B$ decreases and the WH2 distortion function approaches the WH1 one in the limit $B\to0$. Moreover, the gap becomes increasingly greater as $Y$ grows. This is because a larger $Y$ implies a smaller $r$, corresponding to the vicinity of the WH throat. In this region, the strong gravity regime enhances the discrepancies between the two models, especially as $B$ departs from small values. On the other hand, for $Y\to 0$, namely $r\to \infty$, all curves reduce to GR, for which $f(Y)\to0$, due to the asymptotic flatness requirement. 
\begin{figure*}[ht!]
    \centering
    \subfloat[\label{fig:WH1}][WH2]{%
  \includegraphics[trim=0cm 5cm 0cm 2cm,scale=0.29]{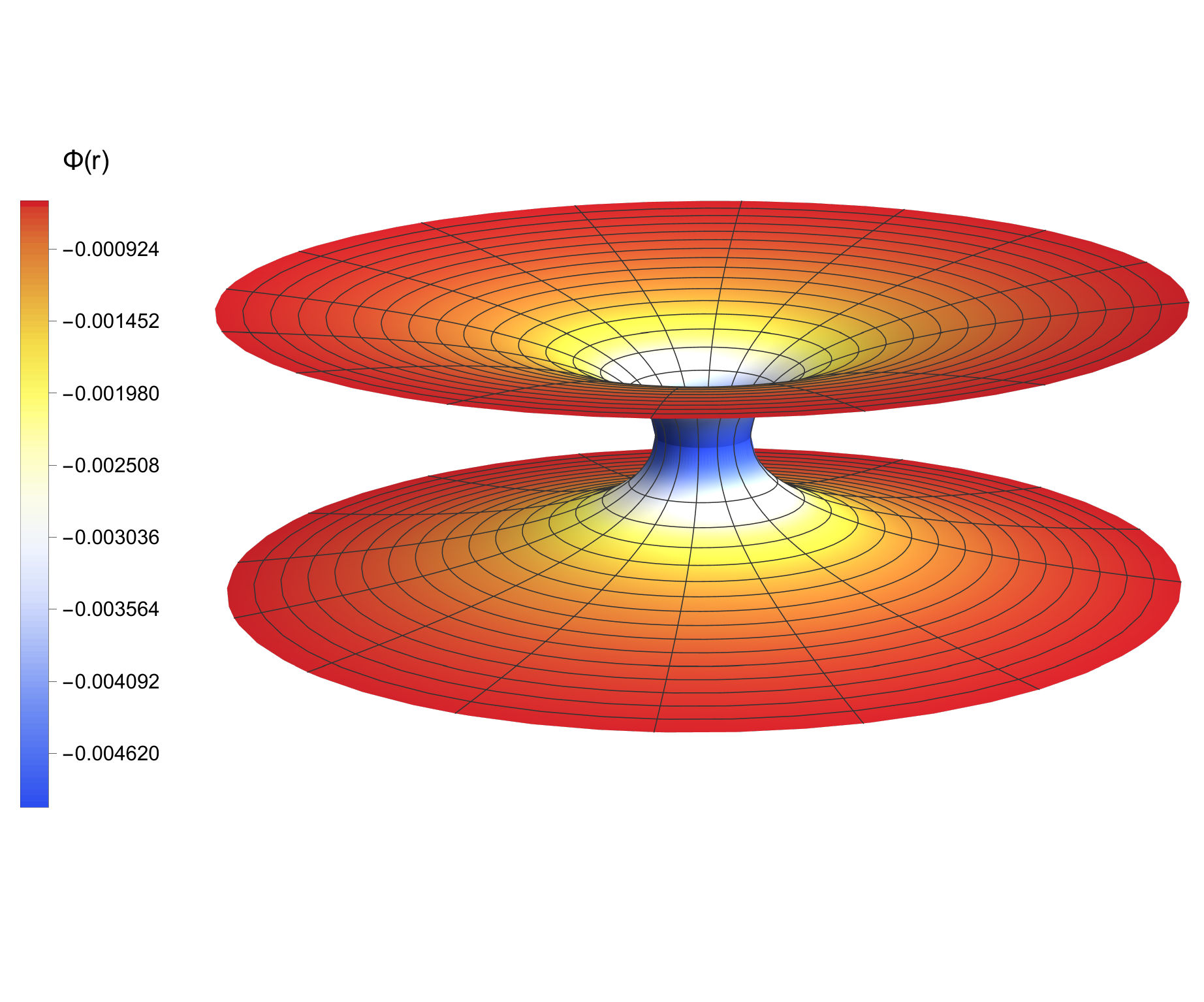}%
}\hfill%
\subfloat[\label{fig:WH2}][WH3]{%
  \includegraphics[trim=0cm 5cm 0cm 2cm,scale=0.29]{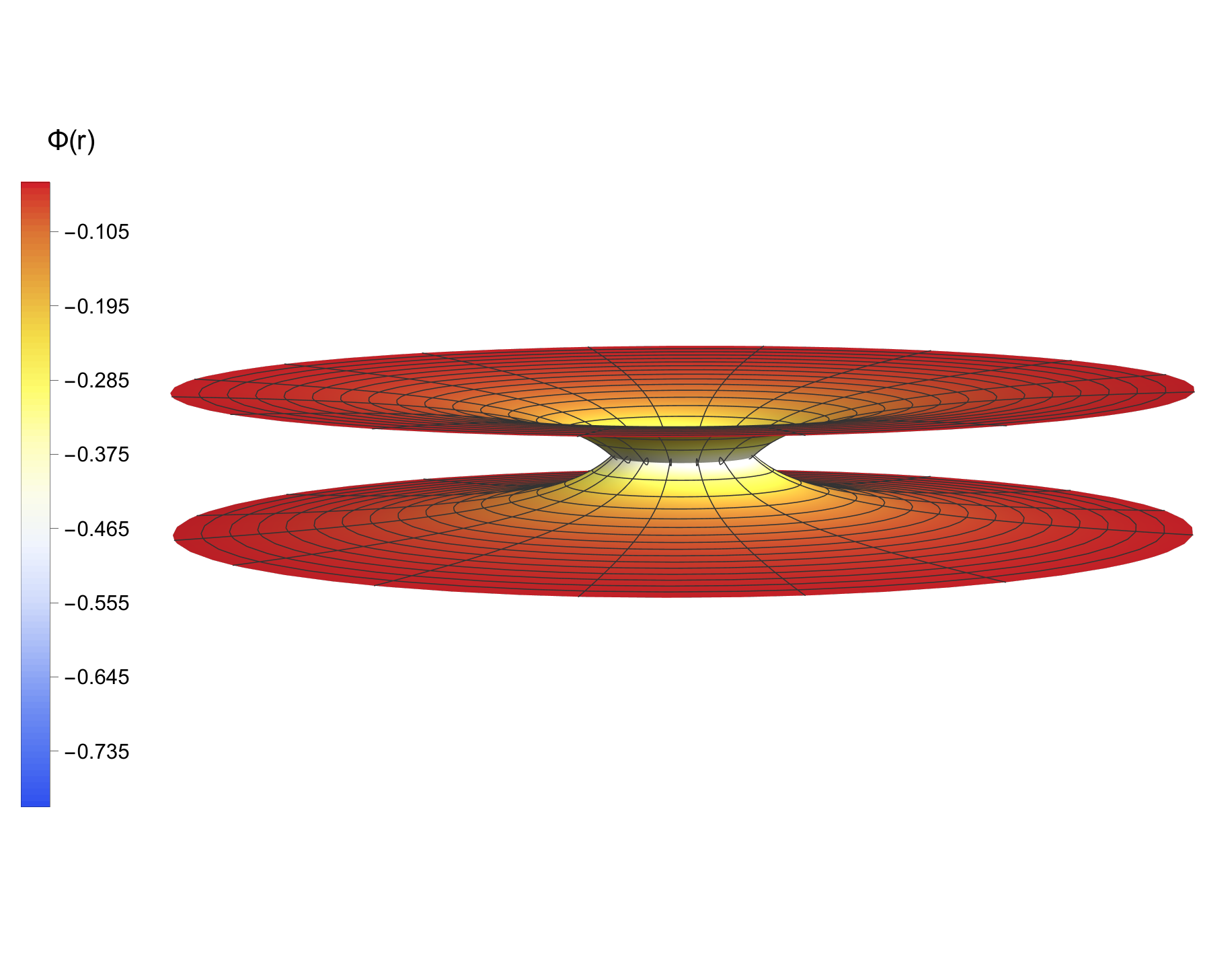}%
}
    \caption{Embedding of the WH2 and WH3 solutions in a three-dimensional Euclidean space for $\theta=\pi/2$, where the shape are provided by $b(r)$, whereas the colors over it represent how the $\Phi(r)$ function varies for $r\in[1,10]$. For the WH2 case, we fix $B=0.01$, whereas for WH3 $B=0.8$. We do not show the WH1 solution as it has a trivial redshift function and its shape looks closely like that of the WH2 solution.}
    \label{fig:Fig1}
\end{figure*}

Furthermore, to quantify the degree of accuracy of the numerical solution compared to the perturbative one, we display in Fig.~\ref{fig:deltaf} the quantity $\Delta f=|f(r)_\text{WH3}-f(r)_\text{WH2}|$ for various settings of $B$. We observe that $\Delta f$ monotonically decreases with $B$, validating thus the perturbative approach. Instead, for larger values of $B$, the WH2 and WH3 solutions increasingly depart from each other. In any case, the absolute error attains its maximum value near the WH throat, due to the singular behavior of $f(Y)$ for $r\to1$\footnote{We underline that this singular behavior is not associated to $b(r)$, (cf. Eq. \eqref{eq:b_sol2}), but to the auxiliary scalar fields.}. The asymptotically flatness guarantees that $\Delta f$ becomes very small ($\lesssim0.1\%$) for $r\gg1$, as all solutions converge to GR. 
In Fig.~\ref{fig:Fig1}, we show the embedding diagrams for the WH2 and WH3 in a three-dimensional Euclidean space, to compare the shape and redshift functions of the two solutions. 

Finally, a summary of the key properties of our WH solutions is provided in Table~\ref{tab:summary}. Specifically, we compare the redshift, shape, and distortion functions for the different cases under study.

\subsection{Energy conditions}
\label{sec:EC}
 An important part of our study must also include the analysis on the energy conditions. In classical GR, there are at least seven different types of energy conditions, which are commonly used to investigate the properties of the WH geometry under investigation \cite{Visser:1995cc}. In the following, we specifically focus on the null energy condition (NEC), typically employed to determine whether the stress-energy tensor, responsible for keeping a WH open and traversable, features ordinary or exotic matter \cite{Morris:1988cz,Visser:1995cc}.

 In our case, we consider a nonlocal gravity theory with vanishing matter stress-energy tensor. However, we can apply the aforementioned analysis by rewriting Eq.~\eqref{eq:FE} in the form of GR with an effective stress-energy tensor, $T^{\rm (eff)}_{\mu\nu}$:
\begin{align} \label{eq:GRmod}
G_{\mu\nu}=\frac{1}{W(r)}T^{\rm (eff)}_{\mu\nu},    
\end{align}
where
\begin{align}
T^{\rm (eff)}_{\mu\nu}&=\left(\nabla_\mu \nabla_\nu-g_{\mu\nu}\Box\right)W-\mathcal{K}_{(\mu\nu)}-\frac{1}{2}g_{\mu\nu}g^{\alpha\beta}\mathcal{K}_{\alpha\beta}\,.
\label{eq:SEeff}
\end{align}
Here, the function $W(r)$ is positive definite (see Eqs. \eqref{eq:SolW1} and \eqref{eq:W_IIcase_sol}), therefore the NEC reads
\begin{equation}\label{eq:NEC}
T^{\rm (eff)}_{\mu\nu}k^\mu k^\nu\ge0\,,    
\end{equation}
for any null vector $k^\mu$. For the metric \eqref{eq:WH-metric}, $k^\mu$ takes the form\footnote{ We assume, without loss of generality, that the motion is confined to the equatorial plane ($\theta=\pi/2$).} \cite{DeFalco:2020afv}
\begin{equation}
k^\mu=\left(-\frac{1}{g_{tt}},\sqrt{\frac{-g_{tt}-r^2b^2}{g_{rr}}},0,\frac{L}{r^2}\right),    
\end{equation}
where $b=L/E$ is the photon impact parameter, while $E$ and $L$ are the conserved energy and angular momentum along the photon trajectory.

\begin{figure*}[ht!]
    \centering
    \hbox{
    \includegraphics[scale=0.29]{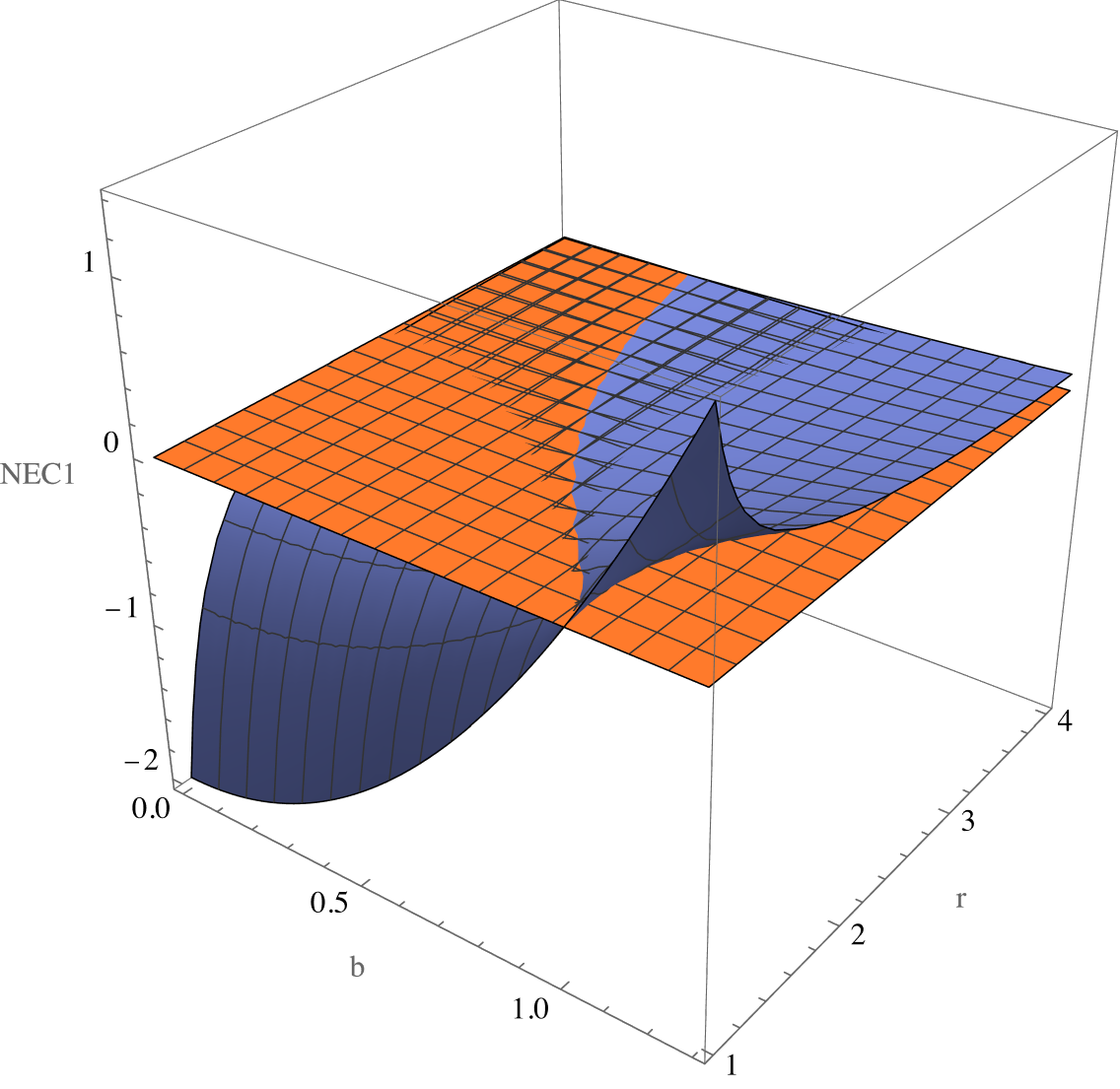}\hspace{0.1cm}
    \includegraphics[scale=0.29]{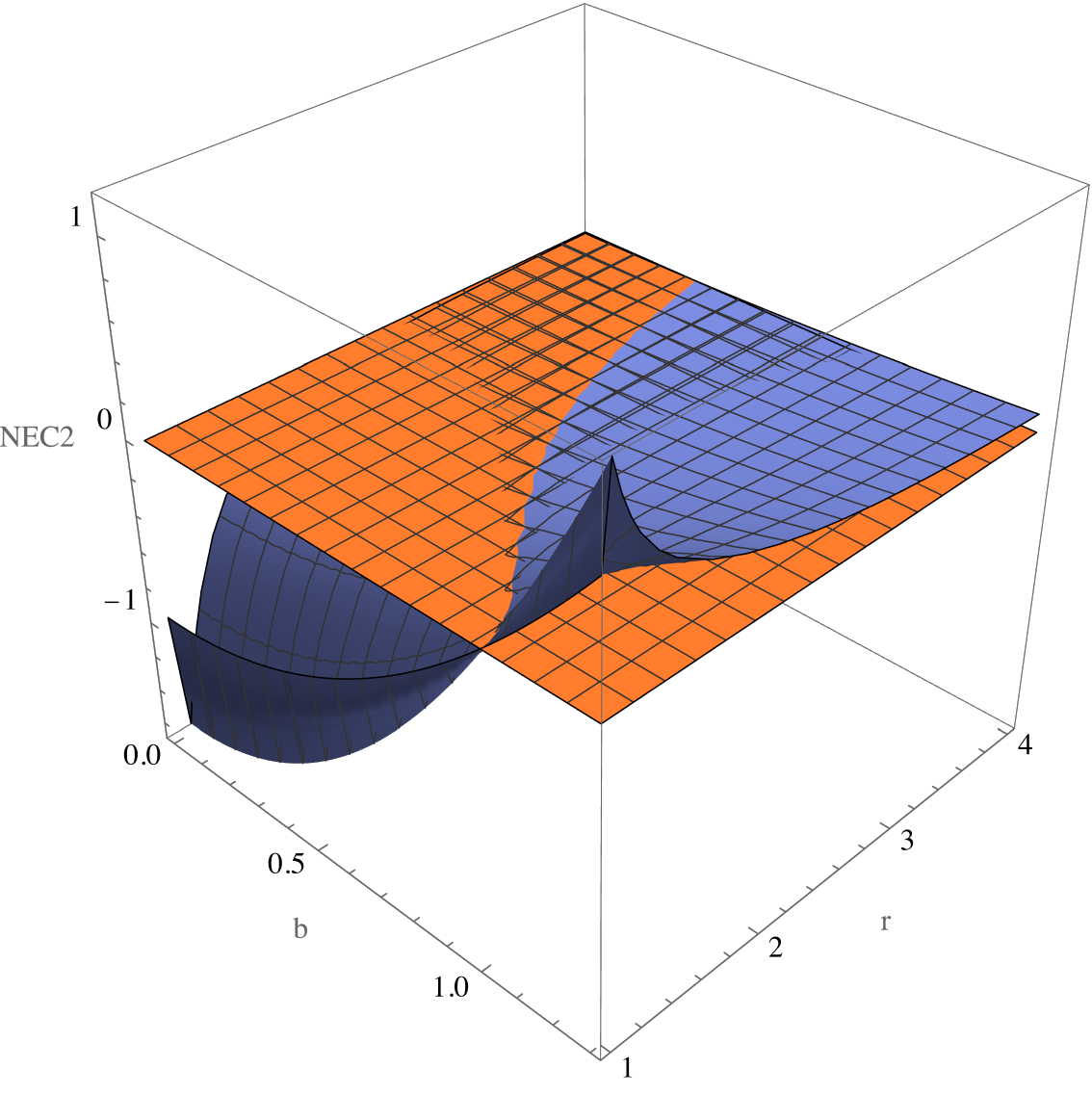}\hspace{0.1cm}
    \includegraphics[scale=0.29]{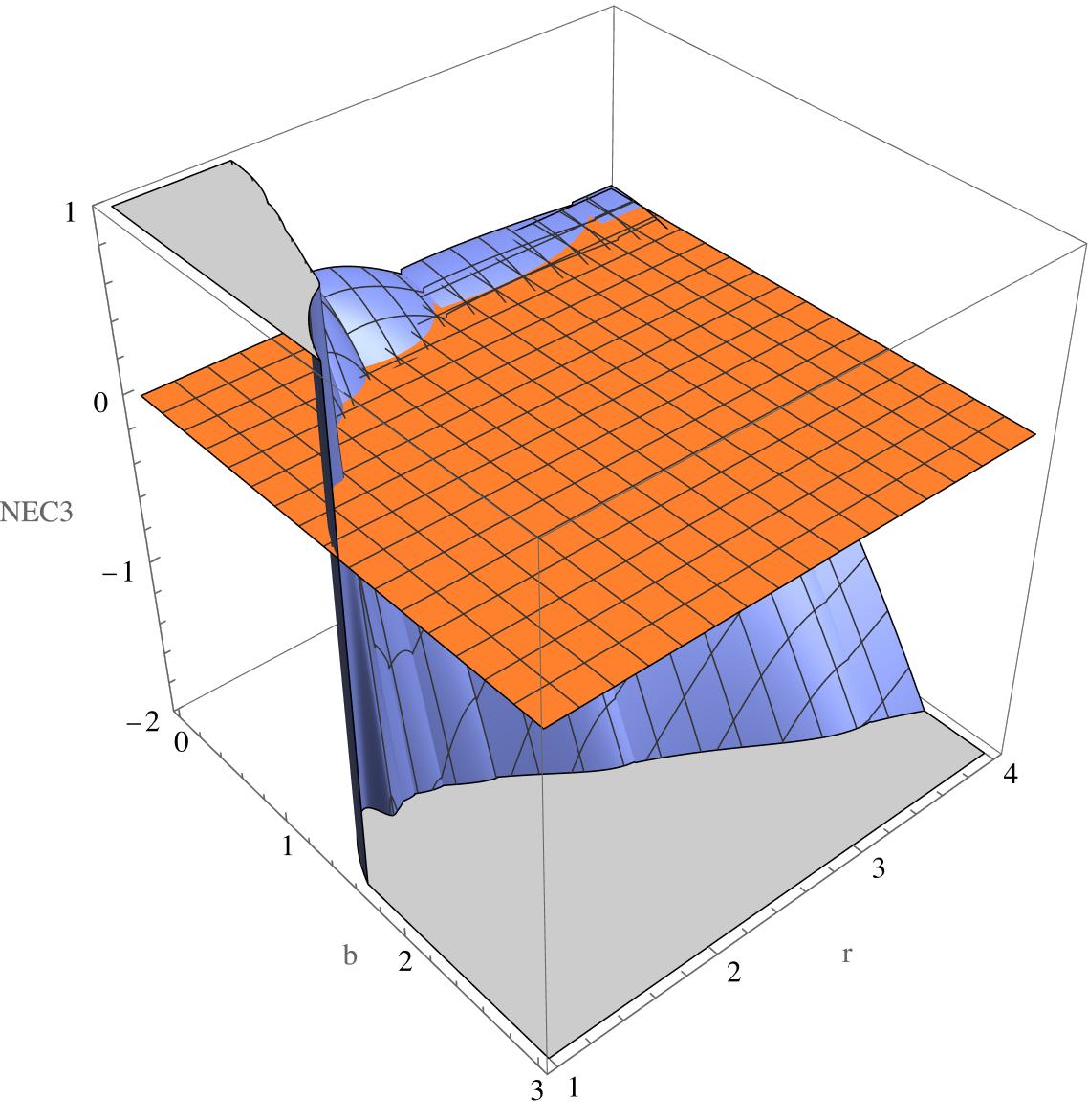}    
    }
    \caption{Three-dimensional plots of Eq.~\eqref{eq:NEC} as functions of $r$ and $b$, where NEC1, NEC2 and NEC3 refer to WH1, WH2 and WH3, respectively. For NEC2, we use $B=0.01$, while $B=0.8$ is chosen. The orange plane is positioned at zero, allowing for the visualization of regions where the NEC takes negative or positive values for the different WH solutions.}
    \label{fig:NEC}
\end{figure*}

Thus, it is essential to examine the sign of Eq.~\eqref{eq:SEeff} for the WH1, WH2, and WH3 solutions. To this end, Fig.~\ref{fig:NEC} portrays three-dimensional representations of Eq.~\eqref{eq:NEC} as functions of $r$ and $b$. Notably, negative values are present in each WH solutions, indicating thus a violation of the NEC. Moreover, for WH2 and WH3, the NEC is consistently violated regardless of the parameter $B$ within its respective range. This finding further supports an intuitive expectation: \emph{nonlocal effects effectively mimic the role of exotic matter in GR due to their quantum nature.}
\begin{table*}[ht!]
\renewcommand{\arraystretch}{3}
\setlength{\tabcolsep}{0.1em}
\centering
\begin{tabular}{|c|c|c|c|}
\hline
 & WH1 & WH2 & WH3 \\ \hline
$\Phi(r)$ & $\Phi_0$ & $-\dfrac{B}{2r}$ & $\dfrac{1}{2}\ln\left(1-\dfrac{B}{r}\right)$\\ \hline
$b(r)$ & $\dfrac{1}{r}$ & $\dfrac{1}{r}+B\left(\dfrac{1-r+\sqrt{r^2-1}-\lambda_1}{r}\right)$ & Numerical \\  \hline
$f(Y)$ & $\dfrac{6}{25} \left\{1-\left[\dfrac{48Y}{\pi ^4}+1\right]^{\frac{5}{4}}+ \ln\left[16 \left(\dfrac{48Y}{\pi ^4}+1\right)^{\frac{5}{4}}\right]\right\}$ & $\dfrac{6}{25} \left\{1-\left[\dfrac{48Y}{\pi ^4}+1\right]^{\frac{5}{4}}+ \ln\left[16 \left(\dfrac{48Y}{\pi ^4}+1\right)^{\frac{5}{4}}\right]\right\}+B f_1(Y)$ & Numerical \\
\hline
\end{tabular}
\caption{Summary of the nonlocal WH solutions found in this work. Here, $f_1(Y)$ is obtained from Eq.~\eqref{eq:f1_IIsol} where $r(Y)$ is given by Eq.~\eqref{eq:r(Y)_sol1}. The functions $b(r)$ and $f(Y)$ corresponding to the WH3 solution cannot be expressed in a closed analytic form and are obtained by numerical integration (see discussion in Sec.~\ref{sec:II-case}).}
\label{tab:summary}
\end{table*}

\section{Wormhole properties}
\label{sec:WH-properties}

 This section is dedicated to examining the fundamental geometric characteristics of the three WH solutions previously derived. This analysis holds significant relevance from an observational standpoint, as the computed quantities correspond to detectable astrophysical phenomena. Consequently, this study serves as the first step toward conducting more sophisticated astrophysical investigations. It is crucial to emphasize that the initial objective is to identify the observational signature of WHs. Once astrophysical data become available, they can be utilized to experimentally assess and validate our WH solutions within the framework of Deser-Woodard nonlocal gravity theory.

In a static and spherically symmetric spacetime, it is interesting to investigate the following quantities \cite{DeFalco:2021klh}:
\begin{itemize}
    \item the photon sphere radius, $r_{\rm ps}$, obtained from
    \begin{equation}\label{eq:rps}
     r\Phi'(r)-1=0\,;   
    \end{equation}
    \item the critical impact parameter, $b_{\rm c}$, which defines the radius of the compact object shadow:
    \begin{equation}\label{eq:bc}
     b_{\rm c}=\frac{r_{\rm ps}}{e^{\Phi(r_{\rm ps})}}\,;  
    \end{equation}
    \item the innermost stable circular orbit (ISCO) radius, $r_{\rm ISCO}$, which is determined by solving the following equation for $r$:
    \begin{equation}\label{eq:rISCO}
     L^2[\Phi'(r)r-1]+\Phi'(r)r^3=0\,,   
    \end{equation}
    where $L$ is the conserved angular momentum along the test particle trajectory. Specifically, $r_{\rm ISCO}$ corresponds to the lowest value of $L$. More generally, the ISCO radius astrophysical represents the inner edge of an accretion disk.
\end{itemize}

We can notice that the solution WH1 does not possess any characteristic radius and therefore neither $b_{\rm c}$. 
For the WH2 solution, we have    
\begin{equation} \label{eq:WH2-values}
r_{\rm ps}=\dfrac{B}{2}\,, \quad b_{\rm c}=\dfrac{e B}{2}\,, \quad r_{\rm ISCO}=B\,,
\end{equation}
while, for the WH3 solution,
\begin{equation} \label{eq:WH3-values}
r_{\rm ps}=\dfrac{3 B}{2}\,,\quad b_{\rm c}=\dfrac{3 \sqrt{3} B}{2}\,,\quad r_{\rm ISCO}=3B\,.
\end{equation}
However, one should bear in mind the flaring out condition \eqref{eq:FOC}, which entails $B>1$ and  $3B/2<1$ for the WH2 and WH3 cases, respectively. These imply that both solutions lack a photon sphere radius and, consequently, also do not have a critical impact parameter. From Eqs.~\eqref{eq:WH2-values} and \eqref{eq:WH3-values}, we observe that for $B\to0$, $1>r_{\rm ps}=0$, which is consistent with our earlier statement regarding the WH1 solution.

Similar arguments can also be applied to $r_{\rm ISCO}$. We note that the WH2 solution does not have this radius, as $1>B>0$, whereas the WH3 case can, in principle, have it, as $2/3>B>1/3$. For $B\to0$, we observe that there is no ISCO radius for the WH1 solution.

Finally, in Fig.~\ref{fig:Fig1}, we display the shape, together with the redshift function, of the WH1 and WH2 solutions embedded in a three-dimensional Euclidean space for $\theta=\pi/2$. The WH1 case is uniformly colored, since $\Phi(r)={\rm const}$. For the selected value of $B=0.01$, the WH2 solution shares a similar shape with the WH3 case, making them almost indistinguishable.

As discussed previously, the obtained WH solutions exhibit highly distinctive characteristics, most notably the absence of a photonsphere and a critical impact parameter. Consequently, from an astrophysical perspective, the WH shadow silhouette coincides precisely with the WH throat. This provides an initial indication of how our WH solutions could be subjected to astrophysical validation. Various methodologies have been proposed in the literature to facilitate the potential detection of WHs. These include, for instance, the study of quasi-normal modes \cite{Konoplya:2016hmd,Cardoso:2016rao,DeSimone:2025sgu}, the analysis of the electromagnetic flux profile from an accretion disk \cite{DeFalco:2020afv}, imaging of accretion disks surrounding WHs \cite{Paul:2019trt,DeFalco:2021klh}, and the examination of epicyclic frequencies \cite{DeFalco:2023twb,DeFalco:2023kqy}. However, investigating the applicability of these approaches to our WH solutions is beyond the scope of the present study.

\section{Discussion and Conclusions}
\label{sec:end}

In this work, we reformulated the field equations of the revised Deser-Woodard nonlocal gravity theory in a proper tetrad frame. This allowed us to reduce the complexity of the equations of motion in vacuum, making them more tractable for analytical, perturbative, and numerical studies. In particular, we focused our attention on a static and spherically symmetric spacetime. We employed a bottom-up procedure, starting by assuming the functional form of the $tt$ metric component and making different ansatz on the functional form of the $g_{rr}$ function. We are then able to solve the set of differential equations for the nonlocal auxiliary fields. Schematically, we first determined the metric tensor, which allowed us to calculate the Ricci scalar. Then, the scalar fields were obtained through a step-by-step method. Therefore, we were able to reconstruct the distortion function of the underlying gravitational theory.
We applied our strategy to search for traversable WHs sustained solely by gravity. Specifically, we demonstrated that exact analytical solutions can be obtained in some simple cases, while for more complex situations, analytical expressions can be found in a perturbative regime. In other cases, numerical routines are required to reconstruct the distortion function. 

 It is worth to remark that, while WHs have been explored in various modified gravity frameworks, including \cite{Menchon:2017qed,Ovgun:2018xys,Javed:2019qyg,Magalhaes:2023har}, our work provides the first systematic investigation of WH solutions within the framework of the revised Deser-Woodard nonlocal gravity model. One of the crucial results of our study is that the obtained classes of WH solutions can be supported without the need for exotic matter. This contrasts with standard GR, where exotic matter is typically required to satisfy the energy conditions. The nonlocal modifications to gravity introduce an effective stress-energy tensor, which allows for a new mechanism of supporting WH geometries. Furthermore, since Deser-Woodard nonlocal gravity has been motivated as an alternative to dark energy, the existence of WHs within this framework may have interesting consequences for cosmology and astrophysics. Also, differently from existing studies that often depend on choosing a specific form of the nonlocal action and making additional assumptions about the scalar fields, our approach enables the derivation of the nonlocal theory within a given spacetime metric without imposing any \emph{a priori} constraints on the gravitational action.

Another important result of our work concerns with the energy conditions. In particular, we demonstrated that the nonlocal effects, encoded in the effective stress-energy tensor, do not satisfy the NEC, thus playing the analogous role of the exotic matter in GR. This result can be understood in light of the quantum nature of nonlocal gravity theories. These findings have significant physical implications in the construction of traversable WHs.

Our results highlight the analytical challenges inherent in nonlocal gravity frameworks, which further motivate the exploration of additional astrophysical solutions. In this respect, the proposed methods hold promise for the discovery of new compact object solutions. In future studies, we aim to build on our developments of new static and spherically symmetric spacetimes, which could contribute to enriching the landscape of exact solutions in nonlocal gravity and provide a deeper understanding of the interplay between modified gravity theories and astrophysical phenomena.

Finally, the methodological advancement presented in this work opens the possibility of extending similar approaches to other nonlocal gravity frameworks, thereby highlighting the versatility and potential impact of our achievements. In addition, our preliminary results may be used to probe nonlocal gravity theories, once accurate observational data are available.

\section*{Acknowledgements}
R.D. acknowledges support from INFN -- Sezione di Roma 1, esperimento Euclid. V.D.F. thanks Gruppo Nazionale di Fisica Matematica of Istituto Nazionale di Alta Matematica, and INFN -- Sezione di Napoli, iniziativa specifica TEONGRAV, for the support.  The authors thank the anonymous referee for insightful comments.

\appendix

\begin{widetext}

\section{Standard metric approach versus tetrad formulation}
\label{sec:full}

The advantages of using the tetrad formalism instead of the standard metric approach can be understood by comparing the original field Eqs.~\eqref {eq:FE} with the ones derived in Eqs.~\eqref{eq:LOC_SUM}. In the former case, the nonvanishing diagonal components read
\begin{subequations}
\begin{align}
&\left(G_{tt}+g_{tt}\Box+\nabla_t\nabla_t\right)W-\frac{1}{2}g_{tt}g^{rr}\mathcal{K}_{rr}=0\,,  \label{eq:original-FE_t}\\
&\left(G_{rr}+g_{rr}\Box+\nabla_r\nabla_r\right)W+\frac{1}{2}\mathcal{K}_{rr}=0 \,,\label{eq:original-FE_r} \\
&\left(G_{\varphi\varphi}+g_{\varphi\varphi}\Box+\nabla_\varphi\nabla_\varphi\right)W-\frac{1}{2}g_{\varphi\varphi}g^{rr}\mathcal{K}_{rr}=0\,. \label{eq:original-FE_phi}
\end{align}
\end{subequations}
These can be reduced to two independent equations by summing Eq.~\eqref{eq:original-FE_t} with Eq.~\eqref{eq:original-FE_phi}, and Eq.~\eqref{eq:original-FE_r} with Eq.~\eqref{eq:original-FE_phi}:
\begin{subequations}
\begin{align}
    &(G_{tt}+G_{\varphi\varphi})W+(\nabla_t\nabla_t+\nabla_\varphi\nabla_\varphi)W+(g_{tt}+g_{\varphi\varphi})\Box W =\frac{1}{2}(g_{tt}+g_{\varphi\varphi})g^{rr}\mathcal{K}_{rr}\,, \label{eq:original_SUM1}\\
    &(G_{rr}+G_{\varphi\varphi})W+(\nabla_r\nabla_r+\nabla_\varphi\nabla_\varphi)W+(g_{rr}+g_{\varphi\varphi})\Box W =\frac{1}{2}(g_{\varphi\varphi}g^{rr}-1)\mathcal{K}_{rr}\,.
    \label{eq:original_SUM2}
\end{align}
\end{subequations}
Then, using the metric \eqref{eq:WH-metric}, we obtain

\begin{subequations}
\begin{align}
& r \left\{ W' \left[ r^2 \left( -b' + 2r \Phi' + 2 \right) + e^{2\Phi} \left( b' - 4 \right) \right] 
    + r \left( e^{2\Phi} - r^2 \right) \left[ X' \left( U' + V X' \right) + V' Y' - 2 W'' \right] \right\} + W \left\{ r^2 \left( r \Phi' + 1 \right) \left( 2r \Phi' - b' \right) \right. \notag \\
    & \left.+ 2r^4 \Phi'' \right\}  
 = b \left\{ r \left( e^{2\Phi} - r^2 \right) \left[ X' \left( U' + V X' \right) + V' Y' - 2 W'' \right] + W' \left( 2r^3 \Phi' + r^2 - 3 e^{2\Phi} \right)\right. \notag \\
    &\left. + r W \left[ r \left( 2r \Phi'' + 2r (\Phi')^2 + \Phi' \right) - 1 \right] \right\}
  \label{eq:original_FE1}\,, \\
    %%%
&b^2 r \left\{ r \left( r \left( X' \left[ U' + V X' \right] + V' Y' - 2 W'' \right) - W' \left[ 2 r \Phi' + 1 \right] \right) - r W \left[ 2 r \Phi'' + 2 r \left( \Phi' \right)^2 + \Phi' \right] + W \right\}\notag\\ 
&- r^2 \left\{ W \left[ r \left( r \Phi' + 1 \right) \left[ 2 r \Phi' - b' \right] + 2 r^3 \Phi'' + 4 \Phi' \right] + W' \left[ 2 \left( \left( r^3 + r \right) \Phi' + r^2 + 2 \right) - r^2 b' \right] + 2 r^3 W''\right.\notag\\
&\left.- r \left( r^2 - 1 \right) \left[ X' \left( U' + V X' \right) + V' Y' \right] \right\} + b \left\{ r \left[ W' \left[ r^2 \left( -b' \right) + 2 \left( 2 r^3 + r \right) \Phi' + 3 r^2 + 4 \right] + 4 r^3 W''\right.\right.\notag\\
&\left.\left. - r \left( 2 r^2 - 1 \right) \left[ X' \left( U' + V X' \right) + V' Y' \right] \right] + W \left[ -r^2 b' \left( r \Phi' + 1 \right) + 4 r^4 \Phi'' - r^2 + r \Phi' \left( 4 r^3 \Phi' + 3 r^2 + 4 \right) + 2 \right] \right\}=0
   . \label{eq:original_FE2}
\end{align}
\end{subequations}
It is evident that Eqs.~\eqref{eq:original_FE1} and \eqref{eq:original_FE2} exhibit significant greater complexity compared to Eq.~\eqref{eq:WH-FE1} and \eqref{eq:WH-FE2}.
This is essentially due to the impossibility to eliminate $\mathcal{K}_{rr}$ from Eqs.~\eqref{eq:original-FE_t}, \eqref{eq:original-FE_r} and \eqref{eq:original-FE_phi}.
Conversely, such elimination can be successfully achieved through Eqs.~\eqref{eq:LOC_SUM1} and \eqref{eq:LOC_SUM2}, demonstrating the remarkable effectiveness of our approach.

\section{First-order perturbations for the WH2 solution}
\label{sec:appendix}
We report here the linear corrections in $B$ related to some scalar fields of Sec.~\ref{sec:II-case}. Specifically, for $V(r)$, we have
\begin{align}
    V_1(r)&=\frac{1}{3200 r \sqrt{r^2-1} \lambda_1^5}\Biggl{\{}64 r \lambda_1^2 \Biggl{[}50v_2 \sqrt{r^2-1}  \lambda_1^6-3 \sqrt{r^2-1} \lambda_1^2-3 \lambda_1 \left(\sqrt{r^2-1} \lambda_3-9 \sqrt{r^2-1}-5\right)\notag\\
    &-15 \left(\sqrt{r^2-1} \lambda_3+1\right)\Biggl{]}+5 r^2 \left(7 \lambda_1^6-42 \lambda_1^4+264 \lambda_1^2-144\right)+5 \left(7 \sqrt{r^2-1} \lambda_1^7-14 \sqrt{r^2-1} \lambda_1^5+168 \sqrt{r^2-1} \lambda_1^3\right.\notag\\
    &\left.+144 \sqrt{r^2-1} \lambda_1-7 \lambda_1^6+42 \lambda_1^4-72 \lambda_1^2+144\right)\Biggl{\}}-\frac{72}{25 \pi ^5 r \sqrt{r^2-1}}\Biggl{[}r \lambda_1^2 \left(6 \sqrt{r^2-1} \lambda_1+5 \sqrt{r^2-1} \lambda_3-5 \lambda_1+5\right)\notag\\
    &-10 r^2 \left(\lambda_1^2-1\right)+5 \left(-2 \sqrt{r^2-1} \lambda_1+\lambda_1^2-2\right)\Biggl{]}-\frac{\pi ^5 \left[7 \text{Si}\left(\frac{\pi }{2}\right)+640 v_2\right]}{20480 \lambda_1^2}+\frac{7 \lambda_1^3 \text{Si}(\lambda_1)}{640}+\frac{57 \pi }{3200 \lambda_1^2}-\frac{7 \pi ^3}{5120 \lambda_1^2}\notag\\
    &-\frac{21}{8 \pi  \lambda_1^2}+\frac{9}{\pi ^3 \lambda_1^2}\,,
    \label{eq:V1_IIsol}
\end{align}
where $\text{Si}(z)\equiv\int_0^z\frac{\sin x}{x} \dd x$\,, and $v_2$ is an integration constant. Moreover, for $U(r)$ we find
\begin{align}
    U_1(r)&=\frac{3 \text{Ci}(\lambda_1)}{5}-\frac{3}{125} \left[25 \text{Ci}\left(\frac{\pi }{2}\right)-3+140 \ln \left(\frac{\pi }{2}\right)\right]+\frac{4 \lambda_1^5}{125 \pi ^5 \left(r^2-1\right)} \Biggl{\{}r^2 \left[25 \pi ^5 v_2+72 (\lambda_1+\lambda_3-1)\right]\notag\\
    &+24 \left(25 \sqrt{r^2-1}-3 \lambda_1-3 \lambda_3+3\right)-25 \pi ^5 v_2\Biggl{\}}+\frac{\lambda_1^4}{800 r} \Biggl{\{}7-\frac{15360}{\pi ^5 (r+1)} \Biggl{[}r^2 (\lambda_1+\lambda_3)+r \left(\lambda_1+\lambda_3-2 \sqrt{r^2-1}\right)\notag\\
    &-\sqrt{r^2-1}\Biggl{]}\Biggl{\}}+\frac{\left(7 \pi ^5 \sqrt{r^2-1}+15360\right) \lambda_1^3}{800 \pi ^5 r}+\frac{3 \sqrt{r^2-1}}{10 r \lambda_1^3}+\frac{\left(23040 \sqrt{r^2-1}-7 \pi ^5\right) \lambda_1^2}{400 \pi ^5 r}+\frac{3}{20 r \sqrt{r^2-1} \lambda_1} \Biggl{[}4 r+14\notag\\
    &-5 r^2+ \sqrt{r^2-1} r (\lambda_1+\lambda_3)\Biggl{]}-\frac{3 \left(7 \pi ^5 \sqrt{r^2-1}-30720\right) \lambda_1}{400 \pi ^5 r}-\frac{3 \left[384 \left(r^2-1\right)+\pi ^5 r\right]}{5 \pi ^5 r \sqrt{r^2-1}}-\frac{3}{10 r \lambda_1^2}+\frac{21}{100 r}\notag\\
    &-\frac{\pi ^5 [1+\ln 32-5 \ln \pi] \left[7 \text{Si}\left(\frac{\pi }{2}\right)+640 v_2\right]}{25600}+\frac{7 \lambda_1^5 \text{Si}(\lambda_1)}{800}-\frac{7 \pi ^5 \text{Si}\left(\frac{\pi }{2}\right) \ln \lambda_1}{5120}-\frac{\pi ^5 v_2 \ln \lambda_1}{8} +\frac{8 \lambda_1^6}{5 \pi ^5}\notag\\
    &-\frac{6}{25} (\lambda_1+\lambda_3) \ln \lambda_1+\frac{84 \ln \lambda_1}{25}+\frac{36 \ln \lambda_1}{\pi ^3}+\frac{57}{800} \pi  \ln \lambda_1-\frac{7 \pi ^3 \ln \lambda_1}{1280}-\frac{21 \ln\lambda_1}{2 \pi }+\frac{3 \lambda_3}{10}+\frac{1152}{5 \pi ^5}\notag\\
    &+\frac{1}{\pi^3}\left[36 \ln \left(\frac{2}{\pi }\right)-\frac{84}{5}\right]-\frac{\pi  \left[139+195 \ln \left(\frac{2}{\pi }\right)\right]}{4000}-\frac{7 \pi ^3 (1+\ln 32-5 \ln \pi )}{6400}-\frac{3 (3+35 \ln 2-35 \ln \pi )}{10 \pi }\,,
\label{eq:U1_IIsol}
\end{align}
where $\text{Ci}(z)\equiv \int_0^z\frac{\cos x}{x} \dd x$. Finally, $f(r)=f_0(r)+Bf_1(r)$ with $f_0(r)$ given by Eq. \eqref{eq:f(r)_sol1} and $f_1(r)$ reads
\begin{align}
f_1(r)&=-\frac{1}{800 \pi ^8}\Biggl{\{}480 \pi ^8 \text{Ci}(\lambda_1)+\frac{1}{5} \lambda_1^5 \Biggl{[}3072 \pi ^3 \left(\frac{25}{\sqrt{r^2-1}}+3 \lambda_1+3 \lambda_3+25\right)-35 \pi ^8 \text{Si}\left(\frac{\pi }{2}\right)-140 \pi ^6-8928 \pi ^4\notag\\
&-268800 \pi ^2+921600\Biggl{]}+\frac{\pi ^3 \lambda_1^4 \left\{15360 \left[\sqrt{r^2-1}-r^2 (\lambda_1+\lambda_3)+r \left(2 \sqrt{r^2-1}-\lambda_1-\lambda_3\right)\right]+7 \pi ^5 (r+1)\right\}}{r (r+1)}\notag\\
&+\frac{\pi ^3 \left(7 \pi ^5 \sqrt{r^2-1}+15360\right) \lambda_1^3}{r}+\frac{240 \pi ^8 \sqrt{r^2-1}}{r \lambda_1^3}+\frac{2 \pi ^3 \left(23040 \sqrt{r^2-1}-7 \pi ^5\right) \lambda_1^2}{r}\notag\\
&+\frac{120 \pi ^8 \left[4 \sqrt{r^2-1} r (\lambda_1+\lambda_3)-5 r^2+4 r+1\right]}{r\lambda_1 \sqrt{r^2-1} }-\frac{6 \pi ^3 \left(7 \pi ^5 \sqrt{r^2-1}-30720\right) \lambda_1}{r}-\frac{480 \left[384 \pi ^3 \left(r^2-1\right)+\pi ^8 r\right]}{r \sqrt{r^2-1}}\notag\\
&-\frac{240 \pi ^8}{r \lambda_1^2}+\frac{168 \pi ^8}{r}+7 \pi ^8 \lambda_1^5 \text{Si}(\lambda_1)+1280 \pi ^3 \lambda_1^6-192 \pi ^8 (\lambda_1+\lambda_3) \ln \lambda_1+336 \pi ^9 \ln \lambda_1+240 \pi ^8 \lambda_3\Biggl{\}}\notag\\
&+\frac{3}{5} \left[1+\text{Ci}\left(\frac{\pi }{2}\right)\right]-\lambda_2+\frac{1}{r}-\frac{6}{5 \pi }+\frac{24}{\pi ^3}-\frac{1152}{5 \pi ^5}+\frac{\pi }{200}  \left[60 \ln \left(\frac{\pi }{2}\right)-7\right].
\label{eq:f1_IIsol}
\end{align}
\end{widetext}

\bibliography{references}

\end{document}